\documentclass[11pt,twoside]{article}
\usepackage[pdftex]{graphicx}
\usepackage{amsmath}
\usepackage{amssymb}
\usepackage{cite}

\newtheorem{theorem}{Theorem}

\newtheorem{example}[theorem]{Example}

\begin{document}


\begin{abstract}
We propose to address in a natural manner, the \textit{modular variable}
concept explicitly in a Schr\"{o}dinger picture. The idea of Modular
Variables was introduced in 1969 by Aharonov, Pendleton and Petersen to
explain certain \textit{non-local} properties of quantum mechanics. Our
approach to this subject is based on Schwinger's finite quantum kinematics
and it's continuous limit.
\end{abstract}



\thispagestyle{plain}

\label{sh}
\begin{center}
{\Large \textbf{%
\begin{tabular}{c} 
An explicit Schr\"{o}dinger picture for Aharonov's \\[-1mm] 
Modular Variable concept%
\end{tabular}
}}
\end{center}

\bigskip

\bigskip

\begin{center}
\textbf{Augusto C\'{e}sar Lobo$^{1*}$ and Clyffe de Assis Ribeiro$^2$ }
\end{center}

\medskip

\begin{center}
\textit{$^1$Physics Department, Federal University of Ouro Preto, Minas Gerais,
 Brazil }
\end{center}

\begin{center}
\textit{$^2$Physics Department, Federal University of Minas Gerais, Minas
Gerais, Brazil } \smallskip
\end{center}

\begin{center}
$^*$Corresponding author e-mail:lobo@iceb.ufop.br\\[0pt]
\end{center}

\medskip

\section{Introduction}

In \cite{aharonov1969modular}, Aharonov and collaborators introduced the
concept of \textit{modular variables} to explain some peculiar non-local
quantum effects as the modular variable exchange between particles and
fields in situations like the well-known Aharonov-Bohm (AB) effect where a
beam of electrons suffers a phase shift from the magnetic field of a
solenoid even without having had any direct contact with the field. This is
contrary to the usual view where the AB phenomenon is explained by a \textit{%
local} interaction between the \textit{particles} and field \textit{%
potentials}, even if the potentials are somewhat unphysical because they are
determined only up to a gauge transformation. More recently, Aharonov and
collaborators have argued for an even wider application of modular variables
to explain certain non-local quantum effects that arise in what may be
considered as paradigmatic phenomena for quantum theory as wave-particle
interference phenomena. Their approach is based on a \textit{Heisenberg
picture}, while we shall address explicitly the same problem within a 
\textit{Schr\"{o}dinger-picture} approach with the help of the mathematical
structure behind Schwinger's Quantum Kinematical Phase Space. We review in
the next section, Schwinger's Finite Quantum Kinematics for those who which
it is unfamiliar and also to fix our own notation. In section III, we carry
out the continuous limit of this finite structure in two distinct ways to
make it easy to understand the transition between our finite analogue
definition of modular variables and the one introduced by Aharonov. In
section IV we discuss the concept of \textit{pseudo-degrees of freedom}, an
idea introduced by one of the authors on how one attributes quantum degrees
of freedom to tensor product spaces that is essential to our proposal of a
finite setting for modular variables \cite{lobo1995number}. In section V we
finally discuss our conception of a modular variable and we present some
examples. We conclude with section VI where we make some additional comments
and set the stage for further work.

\section{Schwinger's Quantum Kinematics}

Let $W^{(N)}$ be a $N$-dimensional quantum space together with an
orthonormal basis $\{ \vert u_{j}\rangle \} $, $(j=0,...N-1)$, that is 
\begin{equation}
\begin{array}{c}
\langle u^{j}|u_{k}\rangle=\delta_{k}^{j}\qquad\mbox{(orthonormality)} \\ 
|u_{j}\rangle\langle u^{j}|=\hat{I}\qquad\mbox{(completeness)}%
\end{array}
\label{completeness and orthonormality}
\end{equation}
where we will use from now on the \textit{sum convention for repeated lower
and upper indices.} These states are \textit{finite position states}. We
follow Schwinger \cite{schwinger2000quantum}\ and define an unitary
translation operator $\hat{V}$ by a \textit{cyclic permutation} over the $\{
\vert u_{j}\rangle \} $ basis:

\begin{equation}
\hat{V}|u_{j}\rangle=|u_{j-1}\rangle  \label{definition of V translation}
\end{equation}
Clearly one has 
\begin{equation}
\hat{V}^{N}=\hat{I}
\end{equation}
so its spectrum is composed by the $N$\textit{-th roots of unity}:%
\begin{equation}
v_{j}=e^{\frac{2\pi i}{N}j}\qquad\mbox{with}\qquad(j=0,...N-1)
\label{def. of vj phase}
\end{equation}

The eigenstates of $\hat{V}$ also form an orthonormal basis $\{ \vert
v_{j}\rangle \} $, the \textit{finite momentum }states. We can repeat the
above procedure defining an unitary translation operator $\hat{U}$\ that
acts upon the momentum basis by the following cyclic permutation:

\begin{equation}
\hat{U}|v_{j}\rangle=|v_{j+1}\rangle  \label{def.of U translation}
\end{equation}

Again it follows that the spectrum of $\hat{U}$ are the $N$\textit{-th roots
of unity }$v_{j}$. It is possible to show that (with an appropriate phase
choice) the eigenstates of $\hat{U}$ are the original position states $%
\{|u_{j}\rangle \}$. Notice that the indices of the complex phase $v_{j}=e^{%
\frac{2\pi i}{N}j}$ and its powers $(v_{j})^{k}=(e^{\frac{2\pi i}{N}%
j})^{k}=e^{\frac{2\pi i}{N}jk}=v_{j}^{k}$ have a double function both as 
\textit{matrix indices} and as \textit{integer powers} of the $v_{j}$\
phase. Actually, because of the built in $MOD$ $N$ structure of the phases,
the indices may be thought as running over the \textit{finite ring }$Z_{N}$
of integers $MOD$ $N$\textit{\ }\cite{ireland1990classical}. In fact, this
matrix is nothing else but the matrix elements of the \textit{Finite Fourier
Transform} in the position (or momentum) basis. In fact, the overlap between
these set of states is given by%
\begin{equation}
\langle u^{k}|v_{j}\rangle =\frac{v_{j}^{k}}{\sqrt{N}}=\frac{1}{\sqrt{N}}e^{%
\frac{2\pi i}{N}kj}  \label{finite plane wave relation}
\end{equation}%
The above relation shows that the finite position and momentum bases form a 
\textit{mutually unbiased basis} (MUB), a concept that has become important
in modern quantum information theory \cite{durt-2010}. Also it is not
difficult to show that

\begin{equation}
\hat{V}^{j}\hat{U}^{k}=v^{jk}\hat{U}^{k}\hat{V}^{j}
\label{finite Heisenberg-Weyl relation}
\end{equation}

The above equation is a kind of \textit{finite} exponentiated analogue of
the usual canonical commutation relations between position and momentum
observables known as the Heisenberg-Weyl relation. As we shall see in the
next section, essentially the same equation holds for the continuum.

\section{The heuristic continuum limit}

We shall implement the \textquotedblleft continuum limit\textquotedblright\
in two different manners. One symmetric and the other non-symmetric between
the position and momentum states.

\subsection{The symmetric approach}

Let $\dim W^{(N)}=N$ be an \textit{odd integer} (with no loss of generality)
so that the index variation is symmetric in relation to \textquotedblleft
zero\textquotedblright: $j=-\frac{N-1}{2},...+\frac{N-1}{2}$. We introduce
the \textquotedblleft scaled\textquotedblright\ variables

\begin{equation}
x_{j}=\Bigg(\frac{2\pi}{N}\Bigg)^{1/2}j\qquad and\qquad y_{k}=\Bigg(\frac{%
2\pi}{N}\Bigg)^{1/2}k
\end{equation}
so that their \textquotedblleft variation\textquotedblright\ is given by

\begin{equation}
\Delta x_{j}=\Delta y_{k}=\Bigg(\frac{2\pi}{N}\Bigg)^{1/2}
\end{equation}
with $\Delta j=\Delta k=1$ as $\Delta x_{j}$ $\rightarrow0$ and $\Delta
y_{k}\rightarrow0$ for $N\rightarrow\infty$. We also \textquotedblleft
scale\textquotedblright\ the position and momentum eigenkets as:

\begin{equation}
\vert q(x_{j})\rangle =\Bigg(\frac{N}{2\pi}\Bigg)^{1/4}\vert u_{j}\rangle
\qquad\mbox{and}\qquad\vert p(y_{k})\rangle =\Bigg(\frac{N}{2\pi}\Bigg)%
^{1/4}\vert v_{k}\rangle  \label{symmetric ket scaling}
\end{equation}
so that we can write the completeness relation as%
\begin{equation*}
 \hat{I}=\vert u_{j}\rangle \langle u^{j}\vert =\vert v_{k}\rangle
\langle v^{k}\vert = \sum^{\frac{N-1}{2}}_{j=-\frac{N-1}{2}}\vert
q(x_{j})\rangle \langle q(x_{j})\vert \Delta x_{j}=\sum^{\frac{N-1}{2}}_{k=-%
\frac{N-1}{2}}\vert p(y_{k})\rangle \langle p(y_{k}\vert \Delta y_{k}
\end{equation*}

One can give a natural heuristic interpretation of the $N\rightarrow\infty$
limit for the above equation as

\begin{equation}
\hat{I}=\int_{-\infty}^{+\infty}\vert q(x)\rangle \langle q(x)\vert
dx=\int_{-\infty}^{+\infty}\vert p(y)\rangle \langle p(y)\vert dy
\label{completeness relation for the continuum}
\end{equation}
and the inner product between these continuous eigenkets may be written as

\begin{equation}
\langle q(x_{j})|p(y_{k})\rangle=\Bigg(\frac{N}{2\pi}\Bigg)^{1/2}\langle
u^{j}|v_{k}\rangle=\frac{1}{\sqrt{2\pi}}v_{k}^{j}=\frac{1}{\sqrt{2\pi}}%
e^{ix_{j}y_{k}}
\end{equation}
so that for $N\rightarrow\infty$ comes:

\begin{equation}
\langle q(x)|p(y)\rangle=\frac{1}{\sqrt{2\pi}}e^{ixy}
\label{plane wave equation for the continuum}
\end{equation}

Note that we use a slightly different notation than usual in the sense that
we distinguish between the \textquotedblleft type\textquotedblright\ of the
eigenvector ($q$ or $p$) from the actual $x$ eigenvalue \cite%
{lobo2002coordinate}. The norm of the $\vert q(x)\rangle $ and $|p(y)\rangle$
states are clearly \textquotedblleft infinite\textquotedblright\ so the
usual \textit{orthonormalization} must be treated with care in a \textit{%
non-usual} manner:

\begin{equation*}
\langle q(x_{j})|q(x_{k})\rangle=\langle p(x_{j})|p(x_{k})\rangle=%
\Bigg\{ 
\begin{array}{c}
0\qquad\quad\mbox{for}\qquad j\neq k \\ 
(\frac{N}{2\pi})^{1/2}\quad\mbox{for}\qquad j=k%
\end{array}
=_{(N\rightarrow\infty)} \Bigg\{ 
\begin{array}{c}
0\quad\mbox{for}\quad x_{j}\neq x_{k} \\ 
\infty\quad\mbox{for}\quad x_{j}=x_{k}%
\end{array}%
\end{equation*}
which is usually written in a more simplified form as

\begin{equation}
\langle q(x)|q(x^{\prime})\rangle=\langle p(x)|p(x^{\prime})\rangle
=\delta(x-x^{\prime})  \label{orthormality for the continuum}
\end{equation}

One may even consider this to be an \textit{heuristic definition} for the 
\textit{Dirac} \textit{Delta }\textquotedblleft function\textquotedblright.
If we are to insert the \textit{Planck constant} $\hbar$ explicitly in
equation (\ref{plane wave equation for the continuum}), one obtains the
well-known \textit{plane wave equation} for the inner product between
position and momentum eigenkets:

\begin{equation}
\langle q(x)|p(y)\rangle=\frac{1}{\sqrt{2\pi\hbar}}e^{ixy/\hbar}.
\label{plane wave with Planck's constant}
\end{equation}
The wave function in the position and momentum basis for an arbitrary state $%
|\psi\rangle\in W^{(\infty)}$ is given respectively by the amplitudes%
\begin{equation}
\begin{array}{c}
\psi_{q}(x)=\langle q(x)|\psi\rangle \\ 
\psi_{p}(y)=\langle p(y)|\psi\rangle%
\end{array}
\label{wave function in position and momentum basis}
\end{equation}
The action of an arbitrary operator $\hat{O}$ upon arbitrary wave functions
are defined as

\begin{equation}
\begin{array}{c}
\hat{O}\psi_{q}(x)=\langle q(x)|\hat{O}|\psi\rangle \\ 
\hat{O}\psi_{p}(y)=\langle p(y)|\hat{O}|\psi\rangle%
\end{array}
\label{action of operators over wave functions}
\end{equation}
The acting of the \textit{translation operators} over the position and
momentum basis are given by

\begin{equation}
\begin{array}{c}
\hat{V}_{\xi }|p(y)\rangle =e^{i\xi y}|p(y)\rangle \\ 
\hat{U}_{\eta }|q(x)\rangle =e^{i\eta x}|q(x)\rangle .%
\end{array}%
\end{equation}%
and

\begin{equation}
\begin{array}{c}
\hat{V}_{\xi }|q(x)\rangle =|q(x-\xi )\rangle \\ 
\hat{U}_{\eta }|p(y)\rangle =|p(y+\eta )\rangle%
\end{array}%
\end{equation}%
where

\begin{equation}
\begin{array}{c}
\hat{V}_{\xi }=e^{i\xi \hat{P}} \\ 
\hat{U}_{\eta }=e^{i\eta \hat{Q}}%
\end{array}
\label{unitary translations operators for the continuum}
\end{equation}%
The \textit{Hermitian infinitesimal generators} of translations are then
identified with the usual \textit{position} and \textit{momentum}
observables $\hat{Q}$ and $\hat{P}$ such that

\begin{equation}
\begin{array}{c}
\hat{Q}|q(x)\rangle=x|q(x)\rangle \\ 
\hat{P}|p(y)\rangle=y|p(y)\rangle%
\end{array}
\label{position and momentum eigenvalue equations}
\end{equation}
obeying the usual Heisenberg canonical commutation relation

\begin{equation}
\lbrack\hat{Q},\hat{P}]=i\hat{I}\qquad\mbox{with}\qquad\hbar=1
\label{Heisenberg commutation relation}
\end{equation}
together with its \textquotedblleft exponentiated\textquotedblright\ form:%
\begin{equation}
\hat{V}_{\xi}\hat{U}_{\eta}=e^{i\xi\eta}\hat{U}_{\eta}\hat{V}
\label{continuous Heisenberg-Weyl relation}
\end{equation}
which is the \textit{continuous analogue} of the finite Heisenberg-Weyl
relation (\ref{finite Heisenberg-Weyl relation}).

\subsection{The non-symmetric continuum limit}

We shall now set a \textit{different scale} for position and momentum as

\begin{equation}
\begin{array}{c}
x_{j}=\frac{\xi}{N}j \\ 
y_{k}=\frac{2\pi}{\xi}k%
\end{array}
\label{non-symmetric scaling of variables}
\end{equation}
so that their \textquotedblleft variation\textquotedblright\ is given
respectively as

\begin{equation}
\begin{array}{c}
\Delta x_{j}=\frac{\xi}{N} \\ 
\Delta y_{k}=\frac{2\pi}{\xi}%
\end{array}%
\end{equation}

Since $\Delta j=\Delta k=1$, then $\Delta x_{j}\rightarrow0$ for $%
N\rightarrow\infty$ (but the same \textbf{not} occurring for $\Delta y_{k}$%
). We also introduce the \textquotedblleft scaled\textquotedblright\ states

\begin{equation}
\begin{array}{c}
\vert q(x_{j})\rangle =\Big(\frac{N}{\xi}\Big)^{1/2}\vert u_{j}\rangle \\ 
\vert p(y_{k})\rangle =\Big(\frac{\xi}{2\pi}\Big)^{1/2}\vert v_{k}\rangle%
\end{array}
\label{non-symmetric state scaling}
\end{equation}
so that

\begin{equation*}
 \hat{I}=\vert u_{j}\rangle \langle u^{j}\vert =\sum^{\frac{\xi}{2}(1-%
\frac {1}{N})}_{j=-\frac{\xi}{2}(1-\frac{1}{N})}\vert q(x_{j})\rangle
\langle q(x_{j})\vert \Delta x_{j}=\vert v_{k}\rangle \langle v^{k}\vert
=\sum_{k=-\frac{N-1}{2}}^{\frac{N-1}{2}}\vert p(y_{k})\rangle \langle
p(y_{k}\vert \Delta y_{k}
\end{equation*}
such that we may write for the \textquotedblleft continuum
limit\textquotedblright

\begin{equation*}
\hat{I}_{\xi}=\int_{-\xi/2}^{+\xi/2}\vert q(x)\rangle \langle q(x)\vert dx=%
\frac{2\pi}{\xi}\sum_{k=-\infty}^{+\infty}\vert p(y_{k})\rangle \langle
p(y_{k}\vert
\end{equation*}
with the position and momentum basis\textit{\ orthonormalization} given by

\begin{equation}
\begin{array}{c}
\langle q(x)|q(x^{\prime})\rangle=\delta(x-x^{\prime}) \\ 
\langle p(y_{j})|p(y_{k})\rangle=\frac{\xi}{2\pi}\delta_{k}^{j}%
\end{array}
\label{non-symmetric orthonormalization}
\end{equation}
where $\hat{I}_{\xi}$ is the \textit{identity operator} over the space of
states that represent \textit{periodic functions} with period $\xi$ and the
inner product between position and momentum eigenstates are essentially
given by the same expression as the one obtained from the symmetric
continuum limit equation (\ref{plane wave equation for the continuum}):

\begin{equation}
\langle q(x)|p(y_{k})\rangle=\frac{1}{\sqrt{2\pi}}e^{ixy_{k}}
\end{equation}
In this case, the position basis is a set with the \textit{power of the
continuum} and the states have \textit{infinite norm} (though with the
eigenvalue spectrum bounded between $-\xi/2$ and $+\xi/2$) while the
momentum states form an \textit{infinite} but \textit{countable} set with 
\textit{finite norm} for each ket. These two different limit procedures
carry essentially the structure of the usual Fourier analysis theory. The
first (symmetric case) embodies the \textit{Fourier transform theory} and
the second is equivalent to the \textit{Fourier Series} expansion for 
\textit{periodic functions}. Note also that we could have \textit{reversed}
the process in the \textit{non-symmetric limit} making the \textit{moment
states} to be \textit{continuous infinite normed} eigenkets while the 
\textit{position states} would be \textit{finite normed} eigenvectors with
eigenvalues taking \textit{discrete values} in a \textit{infinite countable
set}.

\section{Pseudo degrees of freedom}

\qquad If one wants to deal with a system with \textit{two degrees of
freedom,} for the \textit{continuum} limit, it suffices to construct the
tensor product $W=W_{1}\otimes W_{2}$ with the position and momentum states
given respectively by $|q(\vec{x})\rangle=|q(x^{1})\rangle\otimes
|q(x^{2})\rangle$ and $|p(\vec{y})\rangle=|p(y^{1})\rangle\otimes
|p(y^{2})\rangle$ together with \textit{unitary translation} operators $\hat{%
V}_{\vec{\xi}}=\hat{V}_{\xi^{1}}\otimes\hat{V}_{\xi^{2}}\equiv e^{i\vec{\xi}%
\cdot\vec{P}}$ and $\hat {U}_{\vec{\eta}}=\hat{U}_{\eta^{1}}\otimes\hat{U}%
_{\eta^{2}}\equiv e^{i\vec{\eta}\cdot\vec{Q}}$ that act upon the eigenbasis
as

\begin{equation}
\begin{array}{c}
\hat{V}_{\vec{\xi}}|q(\vec{x})\rangle=|q(\vec{x}-\vec{\xi})\rangle \\ 
\hat{U}_{\vec{\eta}}|p(\vec{y})\rangle=|p(\vec{y}+\vec{\eta})\rangle%
\end{array}%
\end{equation}
\begin{figure}[!ht]
\begin{center}
\includegraphics[
height=2.284in,
width=3.307in]%
{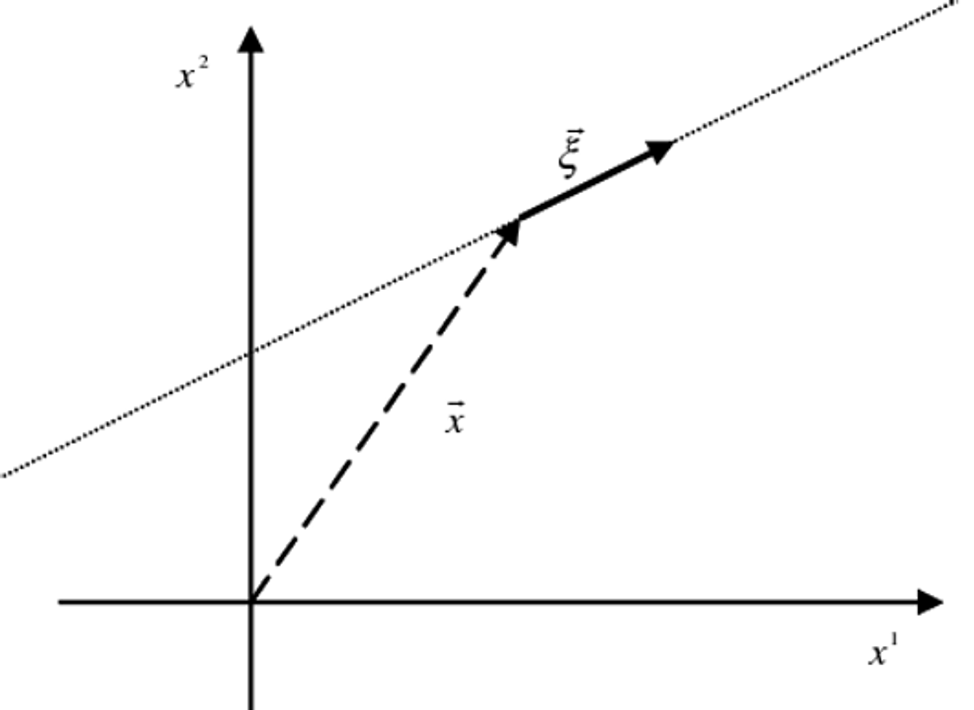}%
\caption{$2$ degrees of freedom}
\label{translacao}
\end{center}
\end{figure}

In the $x^{1}-x^{2}$ plane, one can easily visualize the translations of the
ket $|q(\vec{x})\rangle $ acted repeatedly upon with $\hat{V}_{\vec{\xi}}$
as in figure 1 where the resulting position kets can be represented on a 
\textit{straight line} in the plane that contains point $\vec{x}$ but with
slope given by the $\vec{\xi}$ direction. Of course, to reach an \textit{%
arbitrary point} in the plane, one needs at least\textit{\ two linear
independent} directions. This is precisely what one means when it is \textit{%
said} that the plane is \textit{two-dimensional}. But things for \textit{%
finite} quantum spaces are \textbf{not} quite so simple.\ Let us consider
first a $4$-dimensional system given by the product of two $2$-dimensional
spaces (two qubits) $W^{(4)}=W_{1}^{(2)}\otimes W_{2}^{(2)}$. (It is
important to notice here that one must not confuse the \textit{dimension of} 
\textit{space}, the so called \textit{degree of freedom} with the \textit{%
dimensionality} of the quantum vector spaces). We shall discard in the
following discussion the upper indices that indicate dimensionality to
eliminate excessive notation. So let $\{|u_{0}\rangle ,|u_{1}\rangle \}$ be
the position basis for each individual qubit space so that computational
(unentangled) basis of the tensor product spaces is $\{|u_{0}\rangle \otimes
|u_{0}\rangle ,|u_{0}\rangle \otimes |u_{1}\rangle ,|u_{1}\rangle \otimes
|u_{0}\rangle ,|u_{1}\rangle \otimes |u_{1}\rangle \}$. One may represent
such finite $2$- space as the discrete set formed by the four points
depicted in figure 2. 
\begin{figure}[!ht]
\begin{center}
\includegraphics[
height=2.284in,
width=3.307in
]%
{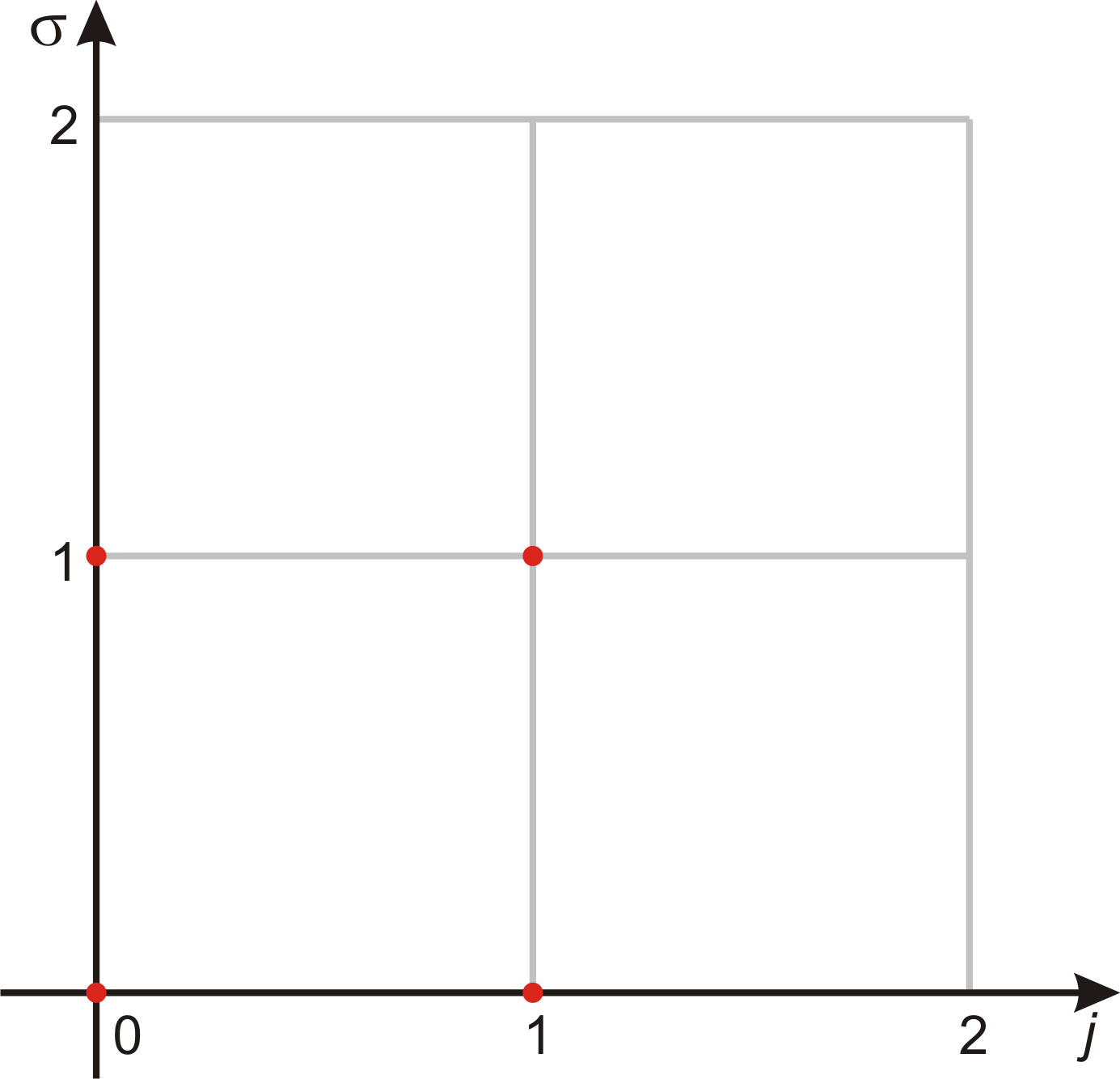}
\caption{Finite 2-space for 2 qubits}
\label{discreto1}
\end{center}
\end{figure}

One may even construct distinct\ "straight lines" in this discrete
two-dimensional space acting upon the computational basis $|u_{j}\rangle
\otimes |u_{\sigma}\rangle $ with the $\hat{V}\otimes \hat{V}$ operator as shown
below:
\begin{figure}[!ht]
\begin{center}
\includegraphics[
height=2.284in,
width=3.307in
]%
{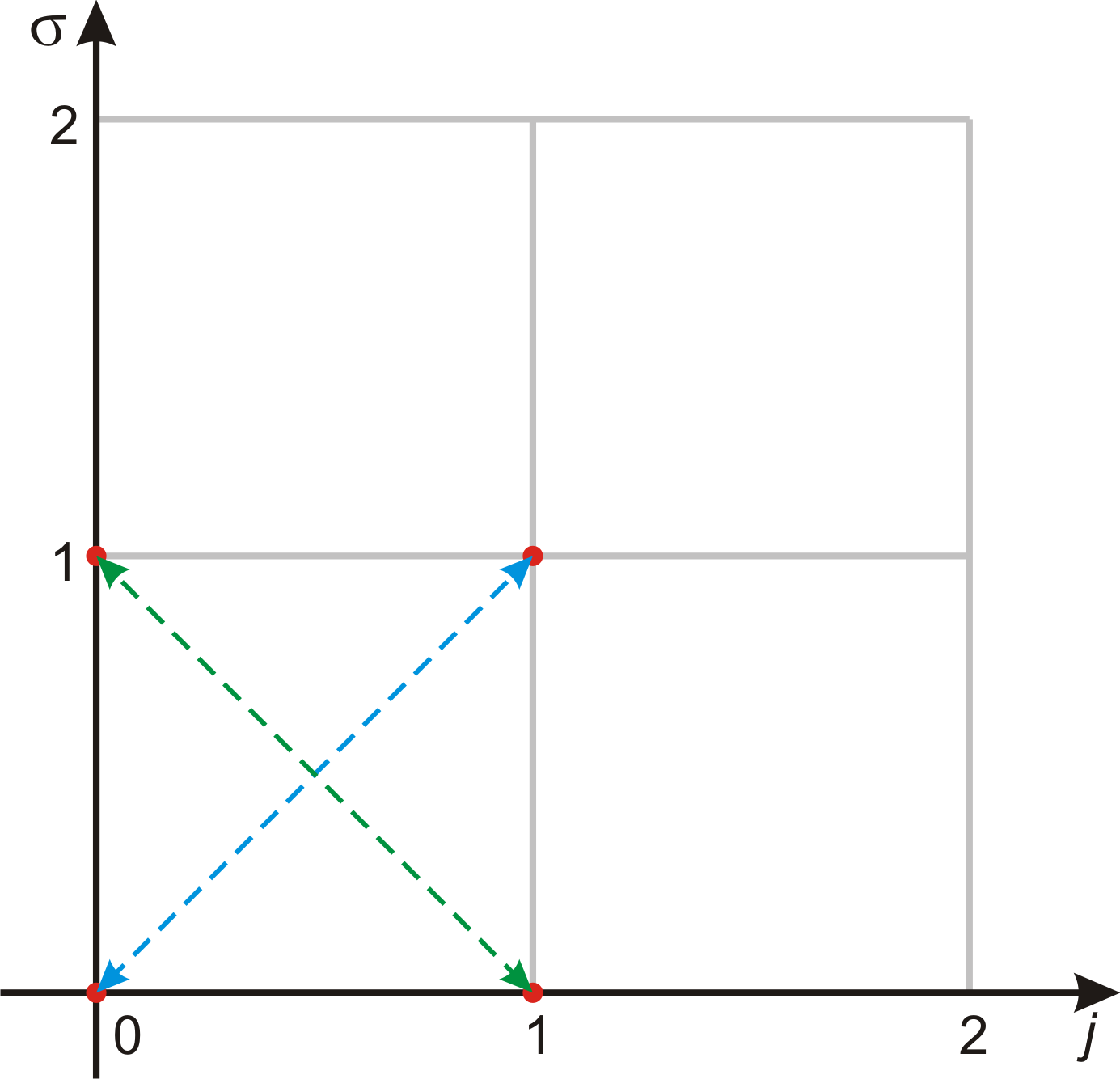}
\caption{Discrete parallel lines $(0,0);(1,1) $ and $(0,1);(1,0)$}
\label{discreto2}
\end{center}
\end{figure}

Each of the two parallel \textquotedblleft straight lines\textquotedblright\
above are \textit{geometric invariants} of the discrete $2$-plane under the
action of $\hat{V}\otimes \hat{V}$. Consider now, a six-dimensional
quantum space $W^{(6)}=W_{1}^{(2)}\otimes W_{2}^{(3)}$ given by the product
of a \textit{qubit} and a \textit{qutrit} space with finite position basis
respectively given by $\{|u_{0}\rangle ,|u_{1}\rangle \}$ and $%
\{|u_{0}\rangle ,|u_{1}\rangle ,|u_{2}\rangle \}$. In this case, the fact
the dimensions of the individual are \textit{coprime} means that the action
of the $\hat{V}\otimes \hat{V}$ operator on the product basis $%
\{|u_{j}\rangle \otimes |u_{\sigma }\rangle \}$ $(j=0,1$ and $\sigma =0,1,2)$
can be identified with the action of $\hat{V}^{(6)}=\hat{V}\otimes \hat{V}$
on the same basis relabeled as $\{|u_{0}\rangle ,|u_{1}\rangle
,|u_{2}\rangle ,|u_{3}\rangle ,|u_{4}\rangle ,|u_{5}\rangle \}$. One can
start with the $|u_{0}\rangle \otimes |u_{0}\rangle $ state and cover the
whole space with \textit{one single} line as shown in figure 4. 
\begin{figure}[!ht]
\begin{center}
\includegraphics[
height=2.3082in,
width=3.3823in
]%
{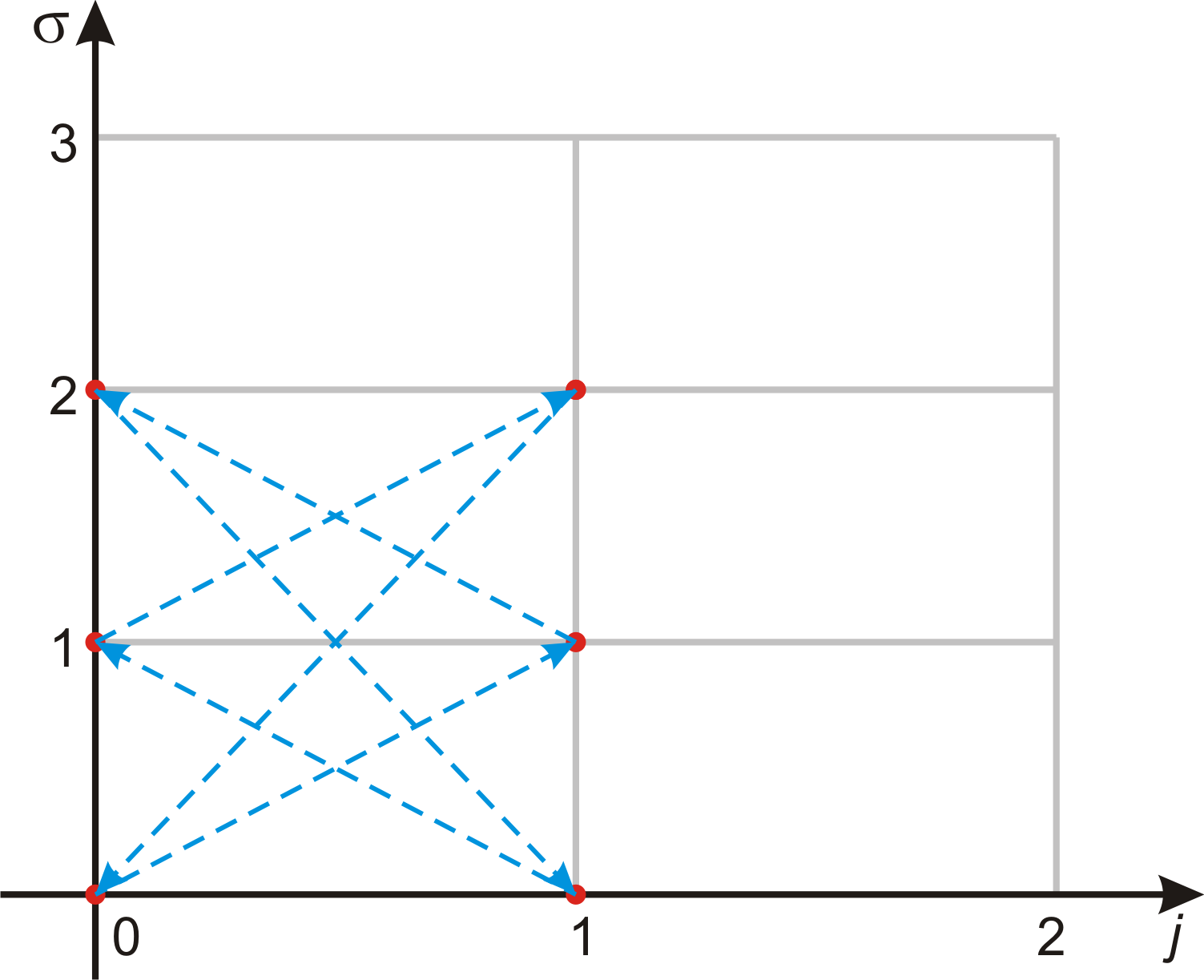}
\caption{$W^{(6)}=W^{(2)}\otimes W^{(3)}$}
\label{1grauliberdade}
\end{center}
\end{figure}


This reduction of \textit{two} degrees of freedom to only \textit{one single}
effective degree of freedom is a general fact for all product spaces when
the dimensions of the factor spaces are \textit{coprime}. In fact we may
state the following theorem (the proof that follows from elementary number
theory can be found in \cite{lobo1995number})

\begin{theorem}
Let $W^{(N)}=W^{(N_{1})}\otimes W^{(N_{2})}$ be the tensor product of two
spaces with $MDC(N_{1},N_{2})=1$ and let each factor space have its own pair
of translation operators $(\hat{U}^{(N_{\alpha})},\hat{V}^{(N_{\alpha})})$
together with a pair of position and momentum basis$\{ \vert
u_{j_{\alpha}}^{(N_{\alpha})}\rangle \} $, $\{ \vert
v_{k_{\alpha}}^{(N_{\alpha})}\rangle \} $, $(j_{\alpha},k_{\alpha}=0,...N_{%
\alpha}-1)$ with $(\alpha=1,2)$ obeying the relations given by (\ref{finite plane wave relation}) and (\ref{finite Heisenberg-Weyl relation}) then if
one defines finite position states $\vert u_{j}\rangle \in W$ by 
\begin{equation}
\vert u_{j}\rangle =(\hat{V}^{t})^{j}\vert u_{0}\rangle
\end{equation}
with 
\begin{equation}
\hat{V}=\hat{V}^{(N_{1})}\otimes\hat{V}^{(N_{2})}
\end{equation}
and%
\begin{equation}
\vert u_{0}\rangle =\vert u_{0}^{(N_{1})}\rangle \otimes\vert
u_{0}^{(N_{2})}\rangle
\end{equation}
then there exists a\textbf{\ single} pair $r_{1}\in Z_{N_{1}}$ and $r_{2}\in
Z_{N_{2}}$ such that the operator

\begin{equation}
\hat{U}=\hat{U}^{r_{1}(N_{1})}\otimes\hat{U}^{r_{2}(N_{2})}
\end{equation}
defines finite momentum states by 
\begin{equation}
\vert v_{k}\rangle =\hat{U}^{k}\vert v_{0}\rangle
\end{equation}
with%
\begin{equation}
\vert v_{0}\rangle =\vert v_{0}^{(N_{1})}\rangle \otimes\vert
v_{0}^{(N_{2})}\rangle
\end{equation}
such that $\hat{V}$ and $\hat{U}$ obey (\ref{finite Heisenberg-Weyl relation}%
) and the finite position and momentum basis $\{|u_{k}\rangle\}\{|v_{k}%
\rangle\}$ obey relation (\ref{finite plane wave relation}).
\end{theorem}

This fact shows that when $MDC(N_{1},N_{2})=1$, then one may say that the
two degrees of freedom are actually \textit{pseudo-degrees} \textit{of
freedom }because one can associate only \textit{one} effective single degree
of freedom to the system. We shall find a physical application of these
pseudo degrees of freedom in terms of Aharonov's \textit{modular variable}
concept in the next section, but first, let us consider some further
examples of the structure contained in the above theorem.

\begin{example}
Let us consider again the six-dimensional case: $N=6$ with $N_{1}=2$ and $%
N_{2}=3$. We may write%
\begin{equation*}
\begin{array}{ccc}
\vert u_{0}^{(6)}\rangle =\vert u_{0}\rangle \otimes\vert u_{0}\rangle ; & 
\vert u_{1}^{(6)}\rangle =\vert u_{1}\rangle \otimes\vert u_{2}\rangle ; & 
\vert u_{2}^{(6)}\rangle =\vert u_{0}\rangle \otimes\vert u_{1}\rangle ; \\ 
\vert u_{3}^{(6)}\rangle =\vert u_{1}\rangle \otimes\vert u_{0}\rangle ; & 
\vert u_{4}^{(6)}\rangle =\vert u_{0}\rangle \otimes\vert u_{2}\rangle ; & 
\vert u_{5}^{(6)}\rangle =\vert u_{1}\rangle \otimes\vert u_{1}\rangle%
\end{array}
\end{equation*}
for the finite position states and compute $r_{1}=3^{-1} (mod \,\,
2)=3^{\phi(2)-1}(mod \,\, 2)=1(mod \,\, 2)$ and $r_{2}=2^{-1}(mod \,\,
3)=2^{\phi(3)-1}(mod \,\, 3)=2(mod \,\, 3)$, so that $\hat{U}=\hat{U}\otimes%
\hat{U}^{2}$ and the momentum states given by:%
\begin{equation*}
\begin{array}{ccc}
\vert v_{0}^{(6)}\rangle =\vert v_{0}\rangle \otimes\vert v_{0}\rangle ; & 
\vert v_{1}^{(6)}\rangle =\vert v_{1}\rangle \otimes\vert v_{2}\rangle ; & 
\vert v_{2}^{(6)}\rangle =\vert v_{0}\rangle \otimes\vert v_{1}\rangle ; \\ 
\vert v_{3}^{(6)}\rangle =\vert v_{1}\rangle \otimes\vert v_{0}\rangle ; & 
\vert v_{4}^{(6)}\rangle =\vert v_{0}\rangle \otimes\vert v_{2}\rangle ; & 
\vert v_{5}^{(6)}\rangle =\vert v_{1}\rangle \otimes\vert v_{1}\rangle .%
\end{array}
\end{equation*}
\end{example}

\begin{example}
As a second example, let us consider $N=15$ with $N_{1}=3$ and $N_{2}=5$. We
can write%
\begin{equation*}
\begin{array}{ccc}
\vert u_{0}^{(15)}\rangle =\vert u_{0}\rangle \otimes\vert u_{0}\rangle ; & 
\vert u_{1}^{(15)}\rangle =\vert u_{2}\rangle \otimes\vert u_{4}\rangle ; & 
\vert u_{2}^{(15)}\rangle =\vert u_{1}\rangle \otimes\vert u_{3}\rangle ; \\ 
\vert u_{3}^{(15)}\rangle =\vert u_{0}\rangle \otimes\vert u_{2}\rangle ; & 
\vert u_{4}^{(15)}\rangle =\vert u_{2}\rangle \otimes\vert u_{1}\rangle ; & 
\vert u_{5}^{(15)}\rangle =\vert u_{1}\rangle \otimes\vert u_{0}\rangle ; \\ 
\vert u_{6}^{(15)}\rangle =\vert u_{0}\rangle \otimes\vert u_{4}\rangle ; & 
\vert u_{7}^{(15)}\rangle =\vert u_{2}\rangle \otimes\vert u_{3}\rangle ; & 
\vert u_{8}^{(15)}\rangle =\vert u_{1}\rangle \otimes\vert u_{2}\rangle ; \\ 
\vert u_{9}^{(15)}\rangle =\vert u_{0}\rangle \otimes\vert u_{1}\rangle ; & 
\vert u_{10}^{(15)}\rangle =\vert u_{2}\rangle \otimes\vert u_{0}\rangle ; & 
\vert u_{11}^{(15)}\rangle =\vert u_{1}\rangle \otimes\vert u_{4}\rangle ;
\\ 
\vert u_{12}^{(15)}\rangle =\vert u_{0}\rangle \otimes\vert u_{3}\rangle ; & 
\vert u_{13}^{(15)}\rangle =\vert u_{2}\rangle \otimes\vert u_{2}\rangle ; & 
\vert u_{14}^{(15)}\rangle =\vert u_{1}\rangle \otimes\vert u_{1}\rangle%
\end{array}
\end{equation*}
for the finite position states and compute $r_{1}=5^{-1}$ $(mod \,\,
3)=5^{\phi(3)-1}(mod \,\, 3)=2(mod \,\, 3)$ and $r_{2}=3^{-1} (mod \,\,
5)=3^{\phi(5)-1}(mod \,\, 5)=2(mod \,\, 5)$, so that $\hat{U}=\hat{U}%
^{2}\otimes\hat{U}^{2}$ and the finite momentum states are%
\begin{equation*}
\begin{array}{ccc}
\vert v_{0}^{(15)}\rangle =\vert v_{0}\rangle \otimes\vert v_{0}\rangle ; & 
\vert v_{1}^{(15)}\rangle =\vert v_{2}\rangle \otimes\vert v_{2}\rangle ; & 
\vert v_{2}^{(15)}\rangle =\vert v_{1}\rangle \otimes\vert v_{4}\rangle ; \\ 
\vert v_{3}^{(15)}\rangle =\vert v_{0}\rangle \otimes\vert v_{1}\rangle ; & 
\vert v_{4}^{(15)}\rangle =\vert v_{2}\rangle \otimes\vert v_{3}\rangle ; & 
\vert v_{5}^{(15)}\rangle =\vert v_{1}\rangle \otimes\vert v_{0}\rangle ; \\ 
\vert v_{6}^{(15)}\rangle =\vert v_{0}\rangle \otimes\vert v_{2}\rangle ; & 
\vert v_{7}^{(15)}\rangle =\vert v_{2}\rangle \otimes\vert v_{4}\rangle ; & 
\vert v_{8}^{(15)}\rangle =\vert v_{1}\rangle \otimes\vert v_{1}\rangle ; \\ 
\vert v_{9}^{(15)}\rangle =\vert v_{0}\rangle \otimes\vert v_{3}\rangle ; & 
\vert v_{10}^{(15)}\rangle =\vert v_{2}\rangle \otimes\vert v_{0}\rangle ; & 
\vert v_{11}^{(15)}\rangle =\vert v_{1}\rangle \otimes\vert v_{2}\rangle ;
\\ 
\vert v_{12}^{(15)}\rangle =\vert v_{0}\rangle \otimes\vert v_{4}\rangle ; & 
\vert v_{13}^{(15)}\rangle =\vert v_{2}\rangle \otimes\vert v_{1}\rangle ; & 
\vert v_{14}^{(15)}\rangle =\vert v_{1}\rangle \otimes\vert v_{3}\rangle .%
\end{array}
\end{equation*}
\end{example}

\section{Modular Variables and pseudo-degrees of freedom}

Consider the \textit{single }degree of freedom associated with the
one-dimensional motion of a particle. Note that because of equation (\ref%
{continuous Heisenberg-Weyl relation}), the $\hat{V}_{L}$ and $\hat{U}_{2\pi
/L}$ operators commute for all values of $L$:%
\begin{equation}
\lbrack \hat{V}_{L},\hat{U}_{2\pi /L}]=0
\label{commuting unitary operators of modular variables}
\end{equation}%
Since these operators are \textit{unitary}, their eigenvalues are
necessarily \textit{complex phases}. Aharonov and collaborators named the
phases of the simultaneous eigenvalues of these pairs of operators as 
\textit{modular variables}. They have shown convincingly the importance of
this object for Quantum Mechanics and, in particular, have argued in favor
of the necessity of the modular variable concept to describe correctly the 
\textit{quantum particle interference} phenomena, a problem that may be
considered as paradigmatic for the general world-vision introduced by
Quantum Physics \cite{aharonov2003quantum}, \cite{tollaksen2010quantum}.
Consider the $n$-slit diffraction experiment where a beam of particles
(electrons for instance) goes through an $n$-slit lattice and strikes a
screen behind it.
\begin{figure}[!ht]
\begin{center}
\includegraphics[
height=2.6013in,
width=4.606in]
{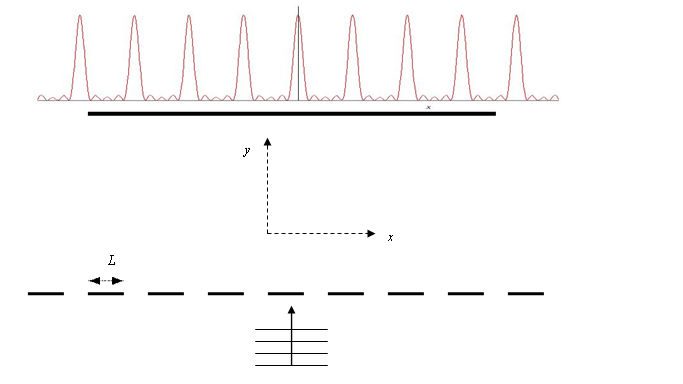}
\caption{n-slit interference experiment}
\label{n-slits}
\end{center}
\end{figure}

The initial state of the incident particle is $|p(\vec{y})\rangle
=|p_{x}(0)\rangle \otimes |p_{y}(\xi )\rangle $. Immediately after the
interaction with the two-slit apparatus, the particle will be in a state $%
|\psi \rangle \otimes |p_{y}(\xi )\rangle $ where $|\psi \rangle $ is a
linear combination of different moment eigenstates in the $x$ direction.
This happens because the particle exchanges modular momentum with the $n$%
-slit screen in the $x$ direction, while leaving unperturbed the particle's $%
y$ degree of freedom. So, from now on, we will concentrate only on the $x$
degree of freedom. Aharonov et al have shown that this state must be an
eigenstate of both the \textit{commuting} unitary translations $\hat{V}%
_{L}=e^{iL\hat{P}}$ and $\hat{U}_{2\pi /L}=e^{i\frac{2\pi }{L}\hat{Q}}$
(equation \ref{commuting unitary operators of modular variables}). This
means that the state $|\psi \rangle $ is simultaneously an eigenstate of
modular momentum and modular position. They presented an example of a phase
space description of such a state as the one given by figure 6 below.
\begin{figure}[!h]
\begin{center}
\includegraphics[
height=2.9897in,
width=4.4183in]%
{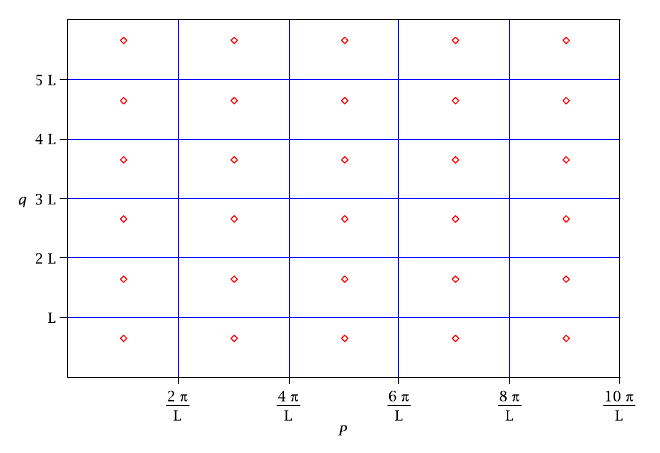}
\caption{state with $q_{\mathrm{mod}}=\frac{2L}{3}$ and $p_{\mathrm{mod}}=\frac{\protect\pi }{L}$}
\label{espaco_fase_aharonov}
\end{center}
\end{figure}
This state has definite values of modular position $q_{%
\mathrm{mod}}=2L/3$ and modular momentum $p_{\mathrm{mod}}=\pi /L$. This
means that the state is represented in each cell by an exact point with
sharp values of $q_{\mathrm{mod}}$ and $p_{\mathrm{mod}}$, but there is a
complete \textit{uncertainty} about \textit{which} cell it belongs to. This
is a basic feature of the modular variable description. We propose a
mathematical description of the finite analogue of this phenomenon in terms
of the \textit{pseudo-degrees of freedom} described in the previous section.
In fact, let $W^{(N)}=W^{(N_{1})}\otimes W^{(N_{2})}$ be a state space for a
quantum mechanical system with $\dim W^{(N)}=N=N_{1}\cdot N_{2}$ and $%
MDC(N_{1},N_{2})=1$. Also each individual factor space carries their finite
position and momentum base states $\{|u_{j_{\alpha }}^{(N_{\alpha })}\rangle
\}$, $\{|v_{k_{\alpha }}^{(N_{\alpha })}\rangle \}$, $(j_{\alpha },k_{\alpha
}=0,...N_{\alpha }-1)$ with $(\alpha =1,2)$ and so we may define finite base
states for $W^{(N)}$ as $|u_{j}^{(N)}\rangle =(\hat{V}%
^{t})^{j}|u_{0}^{(N)}\rangle $ with $|u_{0}^{(N)}\rangle =|u_{0}\rangle
\otimes |u_{0}\rangle $, $\hat{V}^{(N)}=\hat{V}\otimes \hat{V}$ and $%
|v_{k}^{(N)}\rangle =\hat{U}^{k}|v_{0}^{(N)}\rangle $, $|v_{0}^{(N)}\rangle
=|v_{0}\rangle \otimes |v_{0}\rangle $, $\hat{U}=\hat{U}^{r_{1}}\otimes \hat{%
U}^{r_{2}}$ where $r_{1}$ and $r_{2}$\ are given by theorem 7. We can then
offer an interpretation for this \textit{single} degree of freedom of $%
W^{(N)}$ as a degree composed of \textquotedblleft $N_{2}$ \textit{periods}\
of \textit{size} $N_{1}$\textquotedblright\ (or vice-versa). In fact, we may
define the following state of $W^{(N)}$:%
\begin{equation}
|j_{1},k_{2}{}^{(N)}\rangle =|v_{j_{1}}\rangle \otimes |u_{k_{2}}\rangle
\end{equation}%
This state is simultaneously an eigenstate of finite momentum of $%
W^{(N_{1})} $ and finite position of $W^{(N_{2})}$ and clearly represents a
finite analogue of the state represented in figure 4. We may also define the
following operators%
\begin{eqnarray}
\hat{U}^{N_{1}} &=&\hat{I}\otimes \hat{U}^{r_{2}N_{1}}
\label{def. of modular of finite modular operators} \\
\hat{V}^{N_{2}} &=&\hat{V}^{N_{2}}\otimes \hat{I}  \notag
\end{eqnarray}%
which clearly commute:%
\begin{equation}
\lbrack \hat{U}^{N_{1}},\hat{V}^{N_{2}}]=0
\label{finte commuting modular variable operators}
\end{equation}%
which is the finite analogue of equation (\ref{commuting unitary operators
of modular variables}). In fact, it is easy to compute the eigenvalues:%
\begin{eqnarray}
\hat{V}^{N_{2}}|j_{1},k_{2}{}^{(N)}\rangle
&=&v_{j_{1}N_{2}}^{(N)}|j_{1},k_{2}{}^{(N)}\rangle \\
\hat{U}^{N_{1}}|j_{1},k_{2}{}^{(N)}\rangle
&=&v_{k_{2}r_{2}N_{1}}^{(N)}|j_{1},k_{2}{}^{(N)}\rangle  \notag
\end{eqnarray}%
Let us illustrate this with some examples, starting again with the
six-dimensional case:

\begin{example}
$N=6$, $N_{1}=2$ and $N_{2}=3$. we can represent the state $%
|1,2^{(6)}\rangle =|v_{1}\rangle \otimes |u_{2}\rangle $ in the  
\textit{finite phase space} given by figure 7 below.
\begin{figure}[!ht]
\begin{center}
\includegraphics[
height=2.284in,
width=3.307in]%
{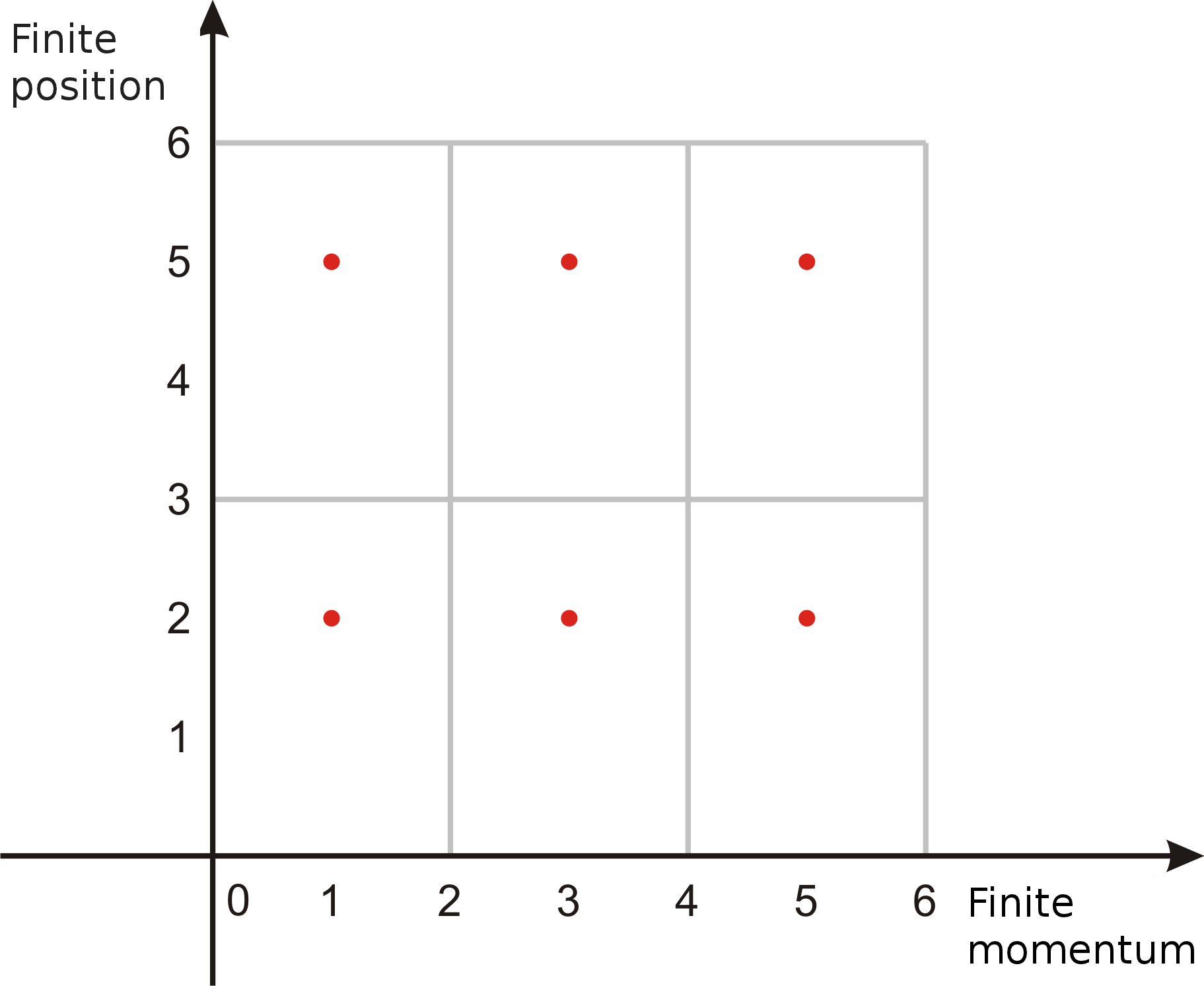}
\caption{$\left\vert 1,2^{(6)}\right\rangle \in W^{(6)}$}
\label{discreto_n6}
\end{center}
\end{figure}

\end{example}

\begin{example}
Let us consider the $N=15$ case with $N_{1}=3$, $N_{2}=5$ and the ket $%
|1,2^{(15)}\rangle =|v_{1}\rangle \otimes |u_{2}\rangle $ represented in the
finite phase space given by figure 8 below. 
\begin{figure}[!hbt]
\begin{center}
\includegraphics[
height=2.284in,
width=3.307in
]%
{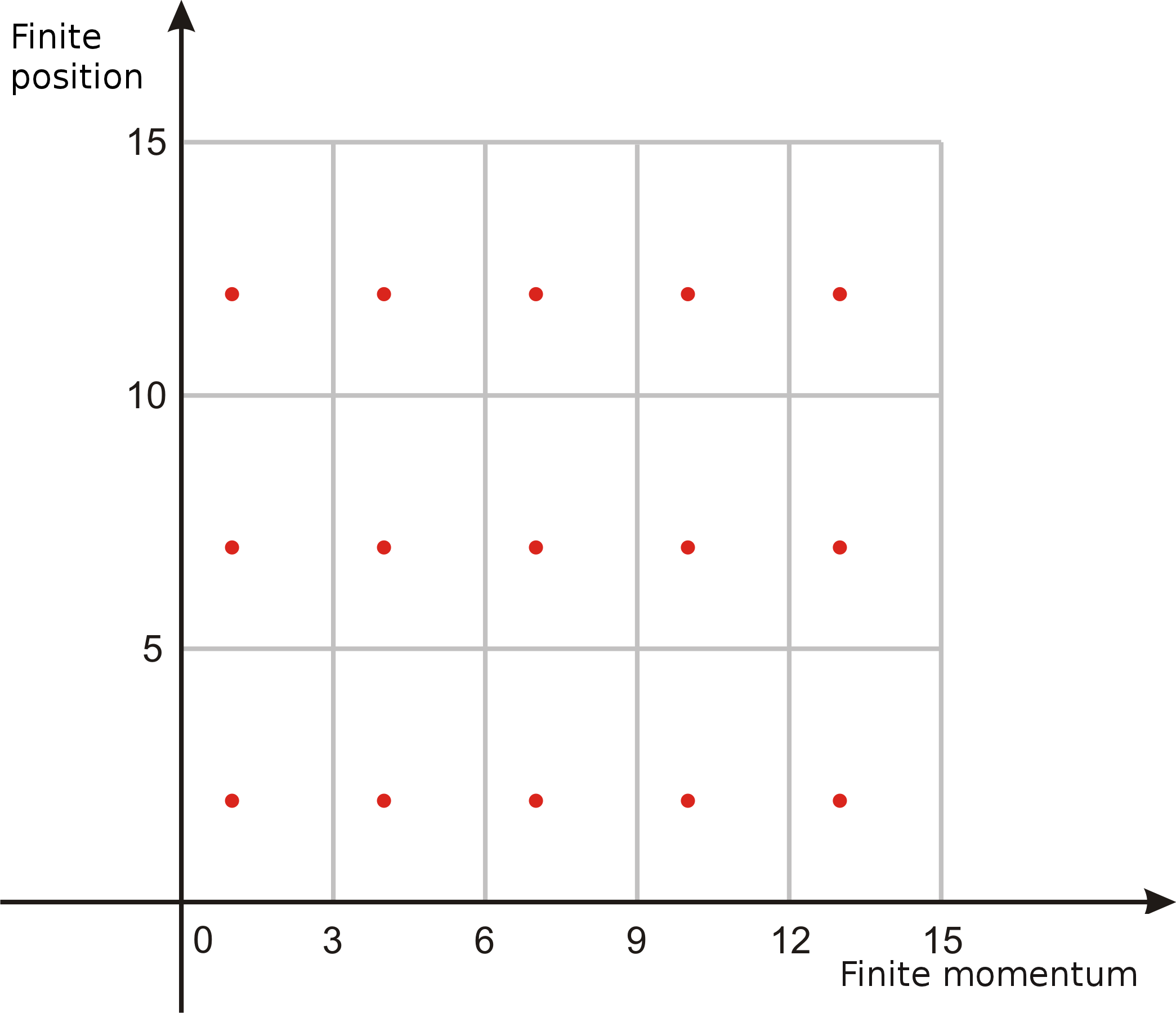}
\caption{$\left\vert 1,2^{(15)}\right\rangle \in W^{(15)}$}
\label{discreto_n15}
\end{center}
\end{figure}

In this case the finite momentum
degree of freedom is composed of $5$ periods of size $3$ and the finite
position degree of freedom is composed of $3$ periods of size $5$.
\end{example}

States with these peculiar mathematical structure have been described 
\textit{independently} by Zak to study systems with \textit{periodic symmetry%
} in quantum mechanics \cite{zak1967finite,zak1968dynamics}. We may call them \textit{Aharonov-Zak states}
(AZ).

This AZ state can be thought as obtained by an \textit{ideal projective
measurement} of modular variables. In fact, the AZ state $|\psi \rangle $
can be obtained starting from $|p_{x}(0)\rangle $ (we only consider the $x$
degree of freedom) and the hamiltonian%
\begin{equation}
\hat{H}(t)=\frac{\hat{P}^{2}}{2m}+V(\hat{Q})\delta (t)\qquad \text{%
with\qquad }V(\hat{Q}+L)=V(\hat{Q})
\end{equation}%
where the particle "hits the screen" at $t=0$ so that the time evolution is
given by $e^{-iV(\hat{Q})}|p_{x}(0)\rangle $. Expanding $e^{-iV(\hat{Q})}$
in a Fourier series gives us 
\begin{equation}
e^{-iV(\hat{Q})}=\sum_{n}c_{n}e^{\frac{2\pi in}{L}\hat{Q}}=\sum_{n}c_{n}\hat{%
U}_{\frac{2\pi }{L}}
\end{equation}%
so that $e^{-iV(\hat{Q})}|p_{x}(0)\rangle $ is \textit{clearly} an
eigenstate both of $\hat{U}_{\frac{2\pi }{L}}$ and $\hat{V}_{L}$. This
mathematical structure is behind the \textit{non-locality} involved in the $%
n $-slit interferometric experiment as thoroughly discussed in \cite{tollaksen2010quantum}. 
The Continuum limit of the AZ state can be constructed in the following manner: the position and momentum
basis of the subsystem $W^{(N_{a})}$ are scaled in the \textit{non-symmetric}
way 
\begin{equation}
|q(x_{j})\rangle =\left( \frac{1}{L}\right) ^{1/2}|u_{j}^{(N_{a})}\rangle
\qquad \text{and\qquad }|p(y_{k})\rangle =\left( \frac{N_{a}L}{2\pi }\right)
^{1/2}|v_{k}^{(N_{a})}\rangle
\end{equation}%
with%
\begin{eqnarray}
x_{j} &=&Lj\qquad \text{so that\qquad }\Delta x_{j}=L \\
y_{k} &=&\frac{2\pi }{LN_{a}}k\qquad \text{so that\qquad }\Delta y_{k}=\frac{%
2\pi }{LN_{a}}  \notag
\end{eqnarray}
and in a similar way (but with the \textit{opposite} construction) we
have for $W^{(N_{b})}$
\begin{equation}
|q(x_{\sigma })\rangle =\left( \frac{N_{b}}{L}\right) ^{1/2}|u_{\sigma
}^{(N_{b})}\rangle \qquad \text{and\qquad }|p(y_{\lambda })\rangle =\left( 
\frac{L}{2\pi }\right) ^{1/2}|v_{\lambda }^{(N_{b})}\rangle 
\end{equation}%
with%
\begin{eqnarray}
x_{\sigma } &=&\frac{L}{N_{b}}\sigma \qquad \text{so that\qquad }\Delta
x_{\sigma }=\frac{L}{N_{b}} \\
y_{\lambda } &=&\frac{2\pi }{L}\lambda \qquad \text{so that\qquad }\Delta
y_{\lambda }=\frac{2\pi }{L}  \nonumber
\end{eqnarray}

And the unitary translation operators:%
\begin{eqnarray}
\hat{V}_{x_{j}}^{\left( N_{a}\right) }|q(x_{k})\rangle
&=&|q(x_{k}-x_{j})\rangle \qquad \hat{V}_{x_{j}}^{\left( N_{a}\right)
}|p(y_{k})\rangle =e^{ix_{j}y_{k}}|p(y_{k})\rangle \\
\hat{U}_{y_{j}}^{\left( N_{a}\right) }|q(x_{k})\rangle
&=&e^{ix_{k}y_{j}}|q(x_{k})\rangle \qquad \hat{U}_{y_{j}}^{\left(
N_{a}\right) }|p(y_{k})\rangle =|p(y_{k}+y_{j})\rangle
\end{eqnarray}%
\begin{eqnarray}
\hat{V}_{x_{\lambda }}^{\left( N_{b}\right) }|q(x_{\sigma })\rangle
&=&|q(x_{\sigma }-x_{\lambda })\rangle \qquad \hat{V}_{x_{\lambda }}^{\left(
N_{b}\right) }|p(y_{\tau })\rangle =e^{ix_{\lambda }y_{\tau }}|p(y_{\tau
})\rangle \\
\hat{U}_{y_{\tau }}^{\left( N_{b}\right) }|q(x_{\sigma })\rangle
&=&e^{ix_{\sigma }y_{\tau }}|q(x_{\sigma })\rangle \qquad \hat{U}%
_{y_{\lambda }}^{\left( N_{b}\right) }|p(y_{\tau })\rangle =|p(y_{\tau
}+y_{\lambda })\rangle
\end{eqnarray}
with identity operators
\begin{eqnarray}
\hat{I}^{\left( N_{a}\right) } &=&\int_{0}^{L}dy|p(y)\rangle \left\langle
p(y)\right\vert =L\sum_{j=0}^{\infty }|q(x_{j})\rangle \left\langle
q(x_{j})\right\vert \\
\hat{I}^{\left( N_{b}\right) } &=&\int_{0}^{L}dx|q(x)\rangle \left\langle
q(x)\right\vert =\frac{2\pi }{L}\sum_{\tau =0}^{\infty }|p(y_{\tau })\rangle
\left\langle p(y_{\tau })\right\vert
\end{eqnarray}%
The continuum limit is then obtained asymptotically with \textit{both} $%
N_{a},N_{b}\rightarrow \infty $.

The finite analogue of the ideal state represented in fig.5 (the Dirac comb
state) can be understood as obtained through an \textit{ideal projective
measurement} carried on by the n-slit apparatus on the incident particle (in
the $x$ degree of freedom):%
\begin{equation}
\left\vert v_{0}\right\rangle \otimes \left\vert v_{0}\right\rangle \overset{%
measurement}{\longrightarrow }\left\vert v_{0}\right\rangle \otimes
\left\vert u_{0}\right\rangle
\end{equation}%
Note that the first subspace is left untouched while the second subspace is
projected to a position eigenstate. This state can be expanded in the
following two ways:%
\begin{equation}
\left\vert v_{0}\right\rangle \otimes \left\vert u_{0}\right\rangle =\frac{1%
}{\sqrt{N}}\sum_{j}\left\vert v_{0}\right\rangle \otimes \left\vert
v_{j}\right\rangle =\frac{1}{\sqrt{N}}\sum_{j}\left\vert u_{j}\right\rangle
\otimes \left\vert u_{0}\right\rangle
\end{equation}%
The first expansion allows us to read the state as a sum over many momentum
states and the second expansion is precisely the \textit{finite analogue} of
the \textit{Dirac comb} state as shown in the figure below: 
\begin{figure}[!hbt]
\begin{center}
\includegraphics[
height=2.284in,
width=3.307in
]%
{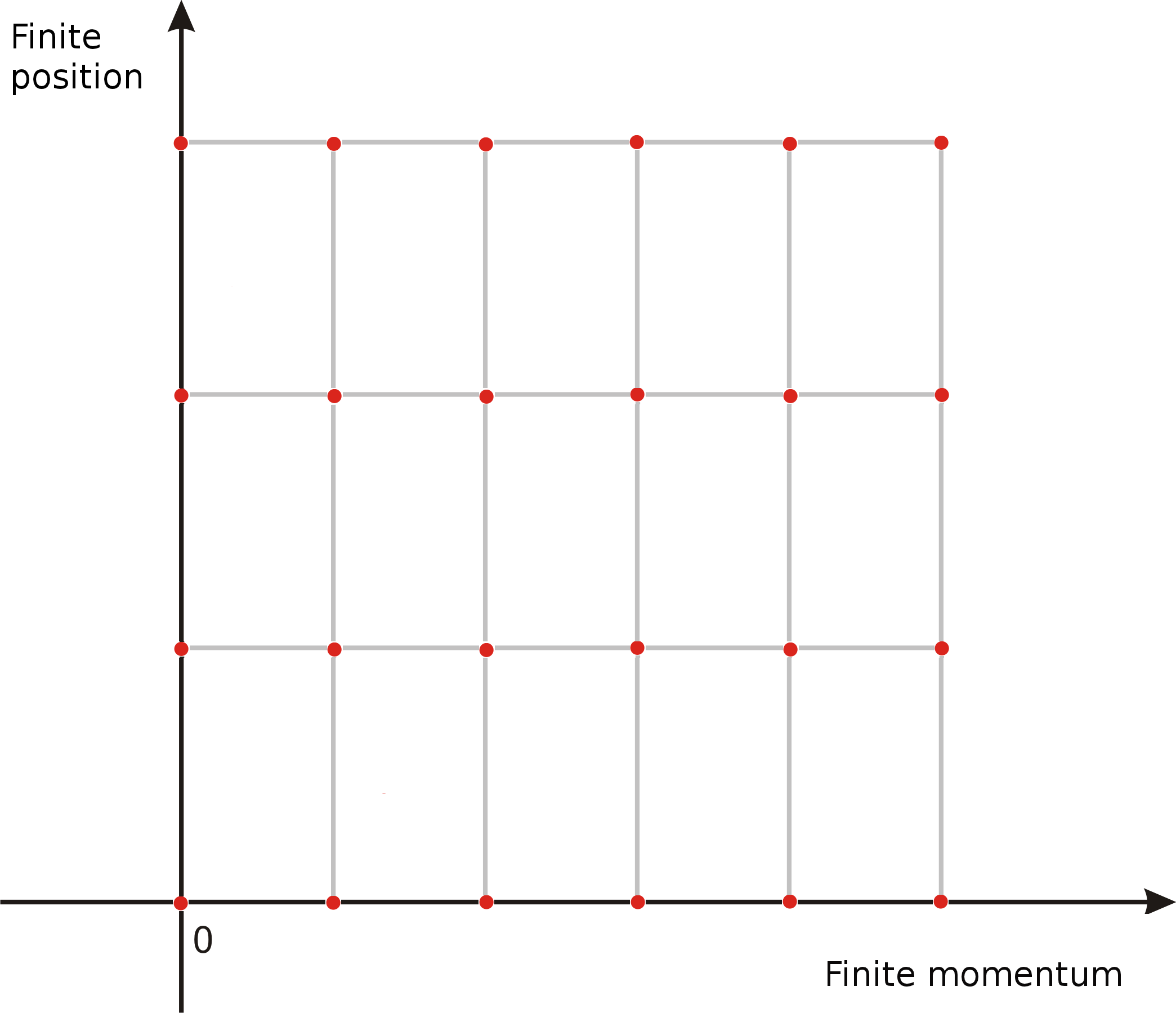}
\caption{Dirac Comb}
\label{discreto_n15}
\end{center}
\end{figure}

\section{Conclusion}

With the advent of Quantum Information Theory, the foundations of Quantum
Mechanics have come back to the main stage of Physics after decades where
this kind of discussion was almost abandoned to what most physicists saw as
a more philosophical kind of concern. Our feeling is that many concepts -
like \textit{modular variables}, the \textit{two-state formalism} and 
\textit{weak values} developed by Aharonov and many important collaborators
are of major importance to help clarify the understanding of Quantum Physics
and Quantum Information and in particular to the comprehension of the still
very elusive phenomena of \textit{quantum non-locality} \cite%
{aharonov1964time}, \cite{aharonov1988result}, \cite{aharonov2002two}. In 
\cite{tollaksen2010quantum}, the authors have addressed the foundational
problem of interference of a particle with a two-slit or multi-slit
apparatus. They offer an explanation of the inherent non-locality in this
experiment in terms of modular variables. How can one conceive that the
electron passing through one slit \textquotedblleft knows\textquotedblright\
that another slit is open or has been closed by a capricious
experimentalist? Their answer is analogous to the one given in \cite%
{aharonov1969modular} for the AB effect. Instead of thinking of this
experiment as a \textit{non-local} Schr\"{o}dinger (physically fictitious)
wave function interacting \textit{locally} with the slits they prescribe an
ontology where \textit{localized} particles interact \textit{non-locally}
(by exchanging modular momentum with the slit screen) in a Heisenberg
representation. And though this non-local interaction \textbf{cannot}
violate causality, with the concept of weak measurements and weak values, it
should be possible to observe in a certain sense, the modular variable
exchange. They also affirm that this non-locality is \textit{dynamical} and
should not be confused with the more \textit{kinematic} kind of non-locality
that happens with EPR like experiments where the non-locality arises because
of the entanglement of the common state of two distant particles. In fact,
they discuss this issue entirely in a Heisenberg-picture framework. We
believe that our analysis implies that, in a certain sense, both kinds of
non-locality arise from the same kind of tensor product space, that can be
carried-out explicitly in a Schr\"{o}dinger-picture, so that these \textit{%
apparently different} kind of non-local quantum phenomena may \textbf{not}
be so unsimilar after all. Recently It has come to our knowledge that the
relation between Schwinger's Finite QM and modular variables has been
noticed before \cite{englert2006periodic}.

We intend to conduct further investigations on this issue and also on the
possibility to extend the above analysis to the \textit{modular energy}
concept \cite{aharonov2003quantum}. We expect difficulties with this last
task because of the lack of symmetry between \textit{time} and \textit{energy%
} in the usual formulation of Quantum Mechanics if compared with the
perfectly symmetric roles played by position and momentum. The power and
flexibility of finite quantum mechanics may be of great help here. The
origin of these ideas may be mostly credited to Schwinger, but further
investigations of the mathematical structure of \textit{finite phase spaces}
have been conducted by many authors since then. See \cite{Galetti1988267}
and \cite{Galetti1992513} for some early results on this subject and also 
\cite{ruzzi2005extended} for some more recent developments.

\section*{Acknowledgments}

Both authors wish to thank the Organizing Committee of the ICSSUR and
Feynman Festival 2011 for the wonderful opportunity and also A. C. Lobo
wishes to acknowledge financial support from Propp-UFOP and NUPEC-Funda\c{c}%
\~{a}o Gorceix and C. A. Ribeiro wishes to acknowledge financial support
from \textit{Conselho Nacional de Desenvolvimento Cient\'{\i}fico e Tecnol%
\'{o}gico} (CNPq).

\bibliographystyle{unsrt}
\bibliography{info_quant}

\end{document}